\documentstyle[12pt]{article}

\begin{document}
\title{
The effect of different baryons impurities }

\author{Yu-Hong Tan$^1$, Xian-Hui Zhong$^2$, Chong-Hai Cai$^2$, Ping-Zhi Ning$^2$\\
$^1$ Theoretical Physics Division, Nankai Institute of
Mathematics,\\ Nankai University,
   Tianjin 300071, P.R.China \\
$^2$ Department of Physics, Nankai University, Tianjin 300071,
P.R.China }
\date{}
\maketitle
\begin{abstract}
We demonstrate the different effect of different baryons
impurities on the static properties of nuclei within the framework
of the relativistic mean-field model. Systematic calculations show
that $\Lambda_c^+$ and $\Lambda_b$ has the same attracting role as
$\Lambda$ hyperon does in lighter hypernuclei. $\Xi^-$ and
$\Xi_c^0$ hyperon has the attracting role only for the protons
distribution, and has a repulsive role for the neutrons
distribution. On the contrary, $\Xi^0$ and $\Xi^+_c$ hyperon
attracts surrounding neutrons and reveals a repulsive force to the
protons. We find that the different effect of different baryons
impurities on the nuclear core
 is due to the different third component of their isospin.
\end{abstract}

\noindent PACS numbers:21.80.+a, 21.10.Dr, 24.10.Jv

\section{Introduction}

 Change of bulk properties of nuclei under the presence of strange
 impurities, like the lambda hyperon ($\Lambda$), is an
 interesting subject in hypernuclear physics. Since a $\Lambda$
 does not suffer from Pauli blocking in $\Lambda$ hypernuclei, it can locate at the center
 of a nucleus, then $\Lambda$ attracts surrounding nucleons
 ($\Lambda$ has the additional attraction provided by - stronger net attraction -
induced attraction) and makes the nucleus
 shrink\cite{prc2001,npa2001}. Recently,
 the experiment KEK-PS E419 has found clear evidence for this shrinkage of
  $^7_\Lambda{\rm Li}$ hypernucleus\cite{ prc2001,npa2001}.

In-medium hyperon interactions have been studied
non-relativistically and relativistically by several groups, e. g.
Hjorth-Jensen et al.\cite{hjo}, Lenske et al.\cite{len}, Ring and
Vretenar.\cite{vre}, Schaffner et al.\cite{96, ann}, Mare${\rm
\breve{s}}$ et al.\cite{mar}. Different from their works, our work
focus on the effect of different baryons impurities on the nuclei.
In the present work, first we will study whether we can obtain
this shrinkage of $\Lambda$ hypernuclei within relativistic
mean-field (RMF) model. After that, it is natural to think whether
other baryons have the attracting role as $\Lambda$ does. In order
to obtain a more profound understanding of the effect of strange
impurities on nuclear core, it is necessary to consider other
impurities, such as $\Sigma$ and $\Xi$, or even heavy flavored
baryons. However, a new experiment at KEK \cite{new7} shows that a
strongly repulsive $\Sigma$-nucleus potential is required to
reproduce the observed spectrum. So, we have reason to believe
that $\Sigma$ hyperon does not have any attracting role and can
not make nucleus shrink. Next in mass are $\Xi^-$ and $\Xi^0$
hyperons. Experimental evidence suggested that the binding energy
of a $\Xi$ hyperon in nuclear matter is negative\cite{6025}.
Therefore we will consider $\Xi$ hypernuclei in this work. In mid
70's and 80's, the theoretical
estimations\cite{r91,r10,pro,r05,npa,r06} predicted a rich
spectrum and a wide range of atomic numbers for charmed and bottom
nuclei. Now, heavy flavored hadrons can be studied at both the
Japan Hadron Facility (JHF) \cite{jhf} and GSI future accelerator
\cite{gsi}, the experimental search for charmed nuclei is becoming
realistic and would be realized. Therefore, We also investigate
the heavy flavored baryons impurities, such as $\Lambda_c^+,
\Lambda_b, \Xi^0_c$ and $\Xi^+_c$. By analogy with $\Sigma$
hyperon, here we do not consider $\Sigma_c$ ($\Sigma^{++}_c$,
$\Sigma^{+}_c$, $\Sigma^0_c$) hypernuclei. Different effect of
different baryons impurities ($\Lambda$, $\Xi^-$, $\Xi^0$,
$\Lambda_c^+$, $\Lambda_b$, $\Xi^0_c$ or $\Xi^+_c$) on nuclear
core is revealed in the present work.

\section{The RMF Model}
To accomplish these, the relativistic mean field model is used.
The RMF model has been used to
 describe nuclear matter, finite nuclei, and hypernuclei successfully.
Here, we start from a Lagrangian density of the form
\begin{equation}
{\cal L} ={\cal L}_{Dirac}+{\cal L}_\sigma +{\cal L}_\omega +{\cal
L}_\rho+{\cal L}_A
\end{equation}
with
\begin{eqnarray}
{\cal L}_{Dirac}&=&\overline\Psi_N(i \gamma^ \mu \partial _\mu
-m_N)\Psi_N+\overline\Psi_{Y}(i \gamma^ \mu \partial_\mu
-m_{Y})\Psi_{Y} \nonumber\\
{\cal L}_\sigma&=&\frac{1}{2}\partial_\mu \sigma \partial^\mu
\sigma -\frac{1}{2}m_\sigma ^2 \sigma^2- g _{\sigma N}
\overline\Psi_N \sigma\Psi_N - g_{\sigma Y}
 \overline{\Psi}_ Y  \sigma\Psi_ Y
     - \frac{1}{3} b {\sigma}^3-\frac{1}{4}c {\sigma}^4 \nonumber \\
 {\cal L}_\omega &=&-\frac{1}{4}F_{\mu\nu}\cdot F^{\mu\nu} + \frac{1}{2}m_\omega ^2\omega_\mu
 \omega^\mu-g_{\omega N} \overline \Psi_N \gamma _\mu\omega^\mu\Psi_N -
 g_{\omega Y}\overline
 \Psi_Y\gamma_\mu\omega^\mu\Psi_Y \\
  {\cal L}_\rho &=&-\frac{1}{4}G_{\mu\nu}\cdot G^{\mu\nu} + \frac{1}{2}m_\rho ^2\rho_\mu
  \rho^\mu-g_{\rho N} \overline \Psi_N \gamma _\mu\rho^\mu\cdot{\bf I}\Psi_N
-g_{\rho Y} \overline \Psi_Y \gamma _\mu\rho^\mu\cdot{\bf I}\Psi_Y  \nonumber\\
    {\cal L}_A &=&-\frac{1}{4}H_{\mu\nu}\cdot H^{\mu\nu} - e
    \overline\Psi_N\gamma_\mu q_N A^\mu\Psi_N \nonumber -e
    \overline\Psi_Y\gamma_\mu q_Y A^\mu\Psi_Y \nonumber
 \end{eqnarray}
 with
 \begin{eqnarray}
   F_{\mu\nu}=\partial_\nu\omega_\mu-\partial_\mu\omega_\nu, \nonumber \\
   G_{\mu\nu}=\partial_\nu\rho_\mu-\partial_\mu\rho_\nu,   \\
H_{\mu\nu}=\partial_\nu A_\mu-\partial_\mu A_\nu,  \nonumber
\end{eqnarray}
where the mesons fields are denoted by $\sigma,\omega_\mu,
\rho_\mu$, and their masses by $m_\sigma, m_\omega, m_\rho$,
respectively. $\Psi_N$ and $\Psi_Y$ are the nucleon and hyperon
fields with corresponding masses $m_N$ and $m_Y$, respectively,
and $Y=\Lambda$, $\Xi^-$, $\Xi^0$, $\Lambda_c^+$, $\Lambda_b$,
$\Xi^0_c$ or $\Xi^+_c$. $A_\mu$ is the electromagnetic fields.
$q_N$ and $q_Y$ are nucleon charge and hyperon charge in the unit
of the proton charge $e$. The Lagrangian for the scalar meson
includes phenomenological non-linear self-interaction, and is
treated in the mean-field and the no-sea approximations
\cite{reinhard}; the contributions of anti(quasi)particles and
quantum fluctuations of mesons fields are thus neglected.

The parametrization (NL-SH) of the nucleonic sector adopted from
Ref.\ \cite{18} is displayed in table \ref{tr}. The center-of mass
correction $E_{c.m.}=-\frac{3}{4}41 A^{-1/3}$ MeV is used for the
RMF forces NL-SH\cite{cm}, where $A$ is the atomic number. First
of all, to check the validity of these parameters,
  we calculate the binding energy per baryon ($-E/A$) and r.m.s. charge
radius ($r_{ch}$), for ordinary nuclei, i.e. the nuclei without
the hyperon. The results are shown in Table \ref{t0}, the
experimental results are also given for comparison. From table
\ref{t0}, it can be found that the properties of finite nuclei can
be well described with this parametrization. For the hyperon
sector, it has been turned out in Ref. \cite{9622,sun5,96,ann}
that the two coupling ratios $g_{\sigma \Lambda}/g_{\sigma N}$ and
$g_{\omega \Lambda}/g_{\omega N}$ of the $\Lambda$ are connected
to the $\Lambda$ potential depth $U_\Lambda$ in nuclear matter by
the relation
\begin{equation}
U_\Lambda=g_{\sigma \Lambda}\sigma^{eq}+g_{\omega
 \Lambda}\omega^{eq}_0=m_N[\frac{m^*_N}{m_N}-1]\frac{g_{\sigma
 \Lambda}}{g_{\sigma N}}+\frac{g_{\omega N}^2}{m_\omega^2}\rho_0\frac{g_{\omega \Lambda}}{g_{\omega N
 }},
\end{equation}
 where $\sigma^{eq}$ and $\omega^{eq}_0$ are the values of
$\sigma$ and $\omega_0$ fields at saturation, and
$m_N^*/m_N=0.597, \rho_0=0.146{\rm fm}^{-3}$ for the set NL-SH.
Hence, for simplicity, similar to Ref. \cite{qmc2}, in an
approximation where the $\omega, \rho$ fields couple only to the
$u$ and $d$ quarks, provided the strange, beauty and charm quarks
in the baryons act as spectators when coupling to the vector
mesons, the coupling constants of hyperons to the vector fields in
the native quark-counting model are obtained as
\begin{eqnarray}
g_{\omega  \Xi^-}=g_{\omega  \Xi^0}=g_{\omega \Xi_c^0}=g_{\omega
\Xi_c^+}=\frac{1}{3}g_{\omega N},\nonumber\\
 g_{\rho \Xi^-}=g_{\rho \Xi^0}= g_{\rho \Xi_c^0}= g_{\rho
\Xi_c^+}=g_{\rho N},\nonumber\\
 g_{\omega  \Lambda}=g_{\omega\Lambda_c^+}=g_{\omega \Lambda_b}=\frac{2}{3}g_{\omega
N}.\nonumber
\end{eqnarray}
$\Lambda$, $\Lambda_c^+$ and $\Lambda_b$ are isoscalar baryons,
and don't couple with $\rho$ meson. Then we fix the scalar
coupling constants to the potential depth of the corresponding
hyperon in normal nuclear matter. Note that $U_Y$ is the
relativistic potential depth. The absolute value of the
nonrelativistic schr$\rm \ddot{o}$dinger equivalent potential
potential depth well will be somewhat smaller$ (10-20)\%$. It is
well known that the potential well depth of $\Lambda$ hyperon in
nuclear matter is about $-30$ MeV, so we use $U_\Lambda=-30$ MeV
 to obtain coupling
constant $g_{\sigma \Lambda}$. However, the experimental data on
$\Xi$ hypernuclei are very little.
 Dover and Gal\cite{dg} analyzed old emulsion data on
 $\Xi^-$ hypernuclei, concluded a nuclear potential well depth of
 $U_\Xi=-21\sim-24$ MeV. Fukuda et al\cite{6024} fitted the very low energy part
 of $\Xi^-$ hypernuclear spectrum in the
 ${\rm ^{12}C(K^-,K^+)X}$ reaction, and estimated the value of $U_\Xi$ to be between
$-16$ and $-20$ MeV. Recently, E885 at the AGS \cite{6025}
indicates a potential depth of $U_\Xi=-14$ MeV or less. Note that
these $\Xi$ potential depth data are estimated based on
Woods-Saxon potentials. Here, we choose $U_{\Xi^-}=U_{\Xi^0}=-16$
MeV to fix $g_{\sigma\Xi}$.
 Because there are no experimental data on
$\Lambda_c^+$, $\Lambda_b$, $\Xi^+_c$ and $\Xi^0_c$ hypernuclei,
the depths of their potential well $U_Y$
 in nuclear matter are not known yet. Ref.\ \cite{r10} estimated the
 $\Lambda_c^+$ nucleus potential was comparable in
 depth to the nucleon-nucleus potential.
 While Ref.\ \cite{pro} suggested $U_{\Lambda_c^+}/U_{\Lambda}\approx 2/3$
and $U_{\Lambda_b}/U_{\Lambda}\approx 1$ within the framework of
the lowest-order Brueckner theory. Ref.\ \cite{npa} reported the
relation between $\Lambda_c^+ N$ potential and $\Lambda N$
potential, roughly $V_{\Lambda_c^+ N }(r)\simeq k V_{\Lambda
N}(r)$, with $k\approx0.8$. Here, we adopt
$U_{\Lambda_c^+}=U_{\Lambda_b}=-30$ MeV, the same as the depth of
$\Lambda$ potential well, to fix the coupling constants of
$\Lambda_c^+$ and $\Lambda_b$ to the scalar meson. Because our
calculations show that $\Xi_c^+$ hypernuclei are very unlikely to
be formed if $|U_{\Xi_c^+}| \leq 14$ MeV, so here
$U_{\Xi^0_c}=U_{\Xi^+_c}=-16$ MeV is chosen. The obtained coupling
constants for hyperons are displayed in table \ref{tr}.

\section{The effect of different baryons impurities}
When a baryon impurity (a baryon different from nucleons) is added
to an ordinary nucleus, the static properties of the nucleus will
be affected.
 In order to observe the universality of
the effect of baryons impurities on nuclear core, an unified RMF
calculation is needed and careful tests should be done. Hence in
our calculations typical hypernuclei between $^7_Y{\rm Li}$ and
$^{209}_Y{\rm Pb}$ are selected, where $Y=\Lambda$, $\Xi^-$,
$\Xi^0$, $\Lambda_c^+$, $\Lambda_b$, $\Xi^0_c$ or $\Xi^+_c$.

Our calculated results for $\Lambda$, $\Xi^-$ and $\Xi^0$
hypernuclei are shown in table \ref{t1} with $U_\Lambda=-30$ and
$U_\Xi=-16$ MeV. The theoretical results for ordinary nuclei are
also given for comparison. In the table, $-E/A$ (in MeV) is the
binding energies per baryon, $r_{ch}$ is the r.m.s. charge radius,
and $r_Y$, $r_n$ and $r_p$ are the calculated r.m.s. radii (in fm)
of hyperon, neutrons and protons distributions, respectively.
Hyperon is at its $1s_{1/2}$ configuration for all the
hypernuclei. From table \ref{t1}, it can be found that for lighter
$\Lambda$ hypernuclei, the size of the core nucleus in a
hypernucleus is smaller than the core nucleus in free space, i.e.,
the values of both $r_n$ and $r_p$ in a hypernucleus are less than
those in the corresponding ordinary nucleus. For instance, the
r.m.s. radius $r_n$ ($r_p$) of neutrons (protons) decreases from
2.32 fm (2.37 fm) in $^6{\rm Li}$ to 2.25 fm (2.29 fm ) in
$^7_\Lambda{\rm Li}$. The attracting role of $\Lambda$ is obtained
in agreement with experiment KEK-PS E419. The attracting role of
$\Lambda$ is also seen in $^9_\Lambda{\rm Be}$ and
$^{13}_\Lambda{\rm C}$ hypernuclei. Above RMF results reveal the
universality of the shrinkage effect for lighter $\Lambda$
hypernuclei. But the situation for $\Xi$ hypernuclei is different.
It is particularly interest to observe a quite different effect
caused by $\Xi$ hyperon impurity.

 From table \ref{t1}, we find:
by adding a $\Xi^-$ hyperon, the r.m.s. radii of the neutrons
become a little larger, while the r.m.s. radii of the protons
become much smaller, comparing with that in the normal nuclei.
Contrary to the $\Xi^-$ hypernuclei, the r.m.s. radii of the
protons become larger and that of the neutrons become smaller in
the $\Xi^0$ hypernuclei. In fact, by calculations, it is found
that the same conclusion is drawn with $-28$ MeV $< U_\Xi < -10$
MeV. The effect of $\Xi^-$ and $\Xi_0$ hyperon on nuclear core is
different from $\Lambda$ hyperon. Note that $\Lambda$, $\Xi^-$ and
$\Xi^0$ are different particles from proton and neutron, they are
all not constrained by the Pauli exclusion. It is obviously that
the common explanation\cite{prc2001} for the $\Lambda$ shrinkage
does not suit the case of $\Xi^-$ and $\Xi^0$. Otherwise, both
$\Lambda$ and $\Xi^0$ hyperon are neutral, hence the origin of the
above difference can not be attributed to the Coulomb potential.
There must be some other source that we don't recognized.

Next, let us to see the effect of heavy flavored baryons
impurities on the nuclear core. The results of $\Lambda_c^+$,
$\Lambda_b$, $\Xi^0_c$ and $\Xi^+_c$ hypernuclei are shown in
table \ref{t2} with $U_{\Lambda_c^+}=U_{\Lambda_b}=-30$ MeV and
$U_{\Xi_c}=-16$ MeV. The results for ordinary nuclei are also
given. The configuration of heavy flavored baryon is $1s_{1/2}$
for all hypernuclei. From table \ref{t2}, it can be seen that both
$r_n$ and $r_p$ become smaller when a $\Lambda_c^+$ or $\Lambda_b$
is added to a lighter nucleus. That is to say $\Lambda_c^+$ and
$\Lambda_b$ have the same attracting role as $\Lambda$ does in
lighter nuclei. While a $\Xi^0_c$ is added to a nucleus, the
situation is the same as adding a $\Xi^-$ hyperon, $r_n$ becomes
lager, and $r_p$ becomes smaller. The effect of adding a $\Xi^+_c$
on nuclear core is the same as a $\Xi^0$ does, $r_p$ becomes
lager, and $r_n$ becomes smaller. Our calculations show that the
effect of $\Xi^0_c$ or $\Xi^+_c$ on nuclear core has the similar
trend as using $-28$ MeV $< U_{\Xi_c} \leq -16$ MeV. From table
\ref{t1} and \ref{t2}, it can be seen that the effect of baryons
impurities on nuclear core is gradually decreasing with increasing
mass number.

In order to understand the different behavior of $\Lambda$ (or
$\Lambda_c^+$, or $\Lambda_b$), $\Xi^-$ (or $\Xi^0_c$) and $\Xi^0$
(or $ \Xi^+_c$) impurities in the nuclei, we make an inspection of
their isospin. $\Lambda$ (or $\Lambda_c^+$, or $\Lambda_b$),
$\Xi^-$ (or $\Xi^0_c$) and $\Xi^0$ (or $\Xi^+_c)$ have different
isospin third component, which may be responsible for their
different behavior. The third component of isospin works through
the coupling of baryon with the $\rho$ mesons in the RMF model. We
may imagine if the couplings of $\rho$ mesons to $\Xi^-$,
$\Xi^0_c$, $\Xi^0$ and $\Xi^+_c$ are omitted from the RMF
calculation, the above mentioned different behavior of $\Xi^-$ (
$\Xi^0_c$) and $\Xi^0$ ($\Xi^+_c$) from $\Lambda$ could disappear.
After eliminating the contribution of the $\rho$ mesons, the RMF
results are shown in table \ref{t3} with $U_\Xi=U_{\Xi_c}=-16$
MeV. From table \ref{t3}, we find the r.m.s. radii of both the
protons and neutrons reduce when adding anyone of these baryons to
the lighter nuclei, which is the same as the situation of adding a
$\Lambda$ hyperon. It is also seen that the effect of baryons
impurities on the heavier nuclei is very little. The nuclear
shrinkage induced by these baryons is obtained in lighter nuclei
when ignoring the contribution of the $\rho$ mesons. The same
conclusion can be obtained with
 $-28$ MeV $< U_\Xi < -10$ MeV or $-28$
MeV $< U_{\Xi_c} < -16$ MeV. While $\Lambda_c^+$, $\Lambda_b$ and
$\Lambda$, $\Xi^0_c$ and $\Xi^-$, $\Xi^+_c$ and $\Xi^0$ have the
same isospin third component, so they have the similar effect on
the nuclear core.

 So, we can conclude that the $\rho$ mesons play an important
role, and the different behavior of $\Lambda$ (or $\Lambda_c^+$,
or $\Lambda_b$), $\Xi^-$ (or $\Xi^0_c$) and $\Xi^0$ (or $
\Xi^+_c$) impurities is due to their different isospin third
component. Although the changes are small, the different response
of $r_p$ and $r_n$ to adding a $\Xi^-$ ($\Xi^0_c$) or $\Xi^0
(\Xi^+_c)$ hyperon may be interesting to know what kind of
properties of the two-body $\Xi N$ ($\Xi_c N$) interaction.
Probably the isospin $T=0$ interaction is attractively large,
while the $T=1$ interaction is repulsive and small. However the
r.m.s. radius is reduced only for one kind of nucleons, but the
r.m.s radius of the other kind of nucleons become larger. It seems
that the nuclei may swell somewhat when adding a $\Xi^-$
($\Xi_c^0$) or $\Xi^0 (\Xi^+_c)$ hyperon. That is very different
from the nuclear shrinkage inducing by a $\Lambda$ in lighter
hypernuclei.

\section{Summary and conclusion}

Within the framework of the RMF theory, we investigate the effect
of different baryons impurities on the nuclear core. The shrinkage
effect induced by a $\Lambda$ hyperon impurity is obtained. It is
found other lighter $\Lambda$ hypernuclei also have this shrinkage
effect besides loosely bound $^6_\Lambda{\rm Li}$. Both
$\Lambda^+_c$ and $\Lambda_b$ have the attracting role as
$\Lambda$ does in lighter hypernuclei.
 We also study the effect of $\Xi$ or $\Xi_c$ hyperon on the
nuclear core. It is found that: by adding a $\Xi^-$ or $\Xi^0_c$
hyperon to the nucleus, $r_n$, the r.m.s. radius of the neutrons
becomes a little larger, while, $r_p$, the r.m.s. radius of the
protons becomes smaller by comparing with that in the core
nucleus. Whereas when adding a $\Xi^0$ or $\Xi^+_c$ hyperon, $r_p$
becomes a little larger and $r_n$ becomes smaller. And this is
very different from the nuclear shrinkage induced by a $\Lambda$
hyperon. We find that the $\rho$ mesons play an important role,
the different effect of $\Lambda$($\Lambda^+_c, \Lambda_b$),
$\Xi^-$ ($\Xi^0_c$), $\Xi^0$ ($\Xi^+_c$) on the nuclear core is
due to their different isospin third component. Although the
changes are small, the different response of $r_p$ and $r_n$ to
adding a $\Xi^-$ ($\Xi^0_c$) or $\Xi^0$ ($\Xi^+_c$) may be
interesting to know what kind of properties of the two-body $\Xi
N$ ($\Xi_c N$) interaction. Probably the isospin $T=0$ interaction
is attractively large, while the $T=1$ interaction is repulsive
and small.

The present work only focuses on the pure $\Lambda$ and $\Xi$
hypernuclei, the coupling between $\Xi N$ and $\Lambda \Lambda$
channels in $\Xi$ hypernuclei isn't taken into consideration. In
addition, we should mention that the coupling constants of
$\Xi^-$, $\Xi^0$, $\Lambda_c^+$, $\Lambda_b$, $\Xi^+_c$ and
$\Xi^0_c$ can not unambiguously be determined, due to be short of
reliable experimental data. In order to get determinate
conclusion, more reliable information are required.

 \hspace{5cm}{ACKNOWLEDGEMENTS}

 This work was supported in part by China postdoctoral science
foundation (2002032169), National Natural Science Foundation of
China (10275037) and China Doctoral
 Programme Foundation of Institution of Higher Education
 (20010055012).

\begin{table}[hb]
 \caption{The coupling constants used in the calculations.
The parametrization (NL-SH) of the nucleonic sector adopted from
Ref.\ \cite{18}, where $m_\sigma =526.059$ MeV, $m_\omega =783$
MeV, $m_\rho=763$ MeV. The vector coupling constants for the
hyperons are taken from the native quark-counting model. The
scalar coupling constants for the hyperons are fixed to the
potential depth of the corresponding hyperon in normal nuclear
matter, $U_\Lambda =U_{\Lambda_c^+}=U_{\Lambda_b}=-30$ MeV,
$U_{\Xi}=U_{\Xi_c}=-16$ MeV.} \label{tr}
 {\small\begin{center}
\begin{tabular}{c|c|c|c|c|c}  \hline
  & $g_{\sigma B}$& $g_{\omega B}$ &$g_{\rho B}$& $b$(fm$^{-1}$)&$c$\\ \hline
 $N$  &10.444 &12.945 &4.383 &-6.9099  &-15.8337 \\ \hline
$\Lambda$&6.4686&8.63&0&0 &0 \\\hline
$\Xi$&3.2619&4.315&4.383&0&0\\\hline\hline
 \end{tabular}
 \end{center}}
\end{table}

\begin{table}[hb]
 \caption{Binding energy per baryon, -$E/A$ (in MeV),
and r.m.s. charge radius $r_{ch}$ (in fm).
 The experimental data of r.m.s. charge radii are taken from \cite{2004}.
  } \label{t0}
 {\small\begin{center}
\begin{tabular}{ccc|cc|ccc|cc}  \hline
  & \multicolumn{2}{c|}{$-E/A$}& \multicolumn{2}{c|} {$r_{ch}$} && \multicolumn{2}{c|}{$-E/A$}& \multicolumn{2}{c} {$r_{ch}$}\\ \hline
 $^AZ$  &RMF &EXP. &RMF &EXP.  &$^AZ$ &RMF &EXP. &RMF &EXP.    \\ \hline
$^6{\rm Li}$&5.67&5.33&2.51&2.54 &$^{16}{\rm O}$&8.04 &7.98 &2.70&2.70 \\
$^{10}{\rm B}$&6.22&6.48&2.46&2.43&$^{40}{\rm Ca}$&8.52& 8.55
&3.46&3.48
\\
$^{12}{\rm C}$&7.47  &7.68&2.46   &2.47         &$^{208}{\rm
Pb}$&7.90& 7.87&5.51&5.50\\\hline\hline
 \end{tabular}
 \end{center}}
\end{table}

\newpage

\begin{table}[hb]
 \caption{Binding energy per baryon, -$E/A$ (in MeV),
 r.m.s. charge radius $r_{ch}$ (in fm),
r.m.s. radii of  hyperon, neutron
 and proton, $r_{Y}$, $r_n$ and $r_p$
 (in fm), respectively.
   The configuration
of hyperons is $1s_{1/2}$ for all hypernclei.
 The results of $\Lambda$ and $\Xi$ hypernuclei are given with $U_\Lambda=-30$ MeV and $U_\Xi=-16$ MeV.
 The experimental data of the ordinary nuclear r.m.s. charge radii are taken from \cite{2004}.
  } \label{t1}
 {\small\begin{center}
 \begin{tabular}{cccccc|ccccccc}  \hline
  $^AZ$&$-E/A$ &$r_{ch}$&$r_{Y}$&$r_n$& $r_p$     &     $^AZ$            &$-E/A$ &$r_{ch}$&$r_{\rm y}$&$r_n$&$r_p$ \\ \hline
 $^6{\rm Li}$&5.67&2.51&&2.32&2.37 &$^{16}{\rm O}$                     &  8.04 &2.70    & &2.55
 &2.58\\\hline
$^7_\Lambda{\rm Li}$&5.63&2.43&2.49&2.25&2.29 &  $^{17}_\Lambda{\rm O}$ &8.33&2.71&2.45&2.55&2.58\\
  $^7_{\Xi^-}{\rm Li}$&5.09&2.41&3.50&2.35 &2.27&  $^{17}_{\Xi^-}{\rm O}$ &8.06&2.68&2.73&2.58&2.55\\
  $^7_{\Xi^0}{\rm Li}$&4.92&2.55&3.90&2.25&2.41 &  $^{17}_{\Xi^0}{\rm O}$ &7.85&2.73&2.89&2.53&2.60\\
  \hline\hline
$^{10}{\rm B}$&6.22&2.46&&2.29&2.32&$^{40}{\rm Ca}$& 8.52  &3.46
&&3.31 &3.36
\\\hline
  $^{11}_\Lambda{\rm B}$&6.63&2.44&2.57&2.28&2.30 &  $^{41}_\Lambda{\rm Ca}$ &8.77&3.46&2.77&3.31&3.36    \\
  $^{11}_{\Xi^-}{\rm B}$&6.14&2.42&2.76&2.32 &2.27&  $^{41}_{\Xi^-}{\rm Ca}$ &8.71&3.44&2.84&3.33&3.34\\
  $^{11}_{\Xi^0}{\rm B}$&5.92&2.49&2.98&2.26&2.35 &  $^{41}_{\Xi^0}{\rm Ca}$ &8.52&3.47&2.98&3.30&3.38\\
  \hline\hline
   $^{12}{\rm C}$  &7.47    &2.46         && 2.30&2.32  &$^{208}{\rm Pb}$&              7.90  &5.51  & &5.71
   &5.45\\\hline
  $^{13}_\Lambda{\rm C}$ &7.90&2.45&2.18&2.28&2.31&   $^{209}_\Lambda{\rm Pb}$ &7.99&5.51&4.13&5.71&5.45    \\
  $^{13}_{\Xi^-}{\rm C}$&7.44&2.42&2.60&2.32 &2.28&  $^{209}_{\Xi^-}{\rm Pb}$ &8.00&5.50&3.72&5.72&5.44\\
  $^{13}_{\Xi^0}{\rm C}$&7.21&2.48&2.77&2.27&2.34 &  $^{209}_{\Xi^0}{\rm Pb}$ &7.95&5.51&4.10&5.70&5.45\\
  \hline\hline

\end{tabular}
\end{center}}
\end{table}

\newpage

\begin{table}
\caption{Binding energy per baryon, -$E/A$ (in MeV), r.m.s. charge
radius $r_{ch}$(those of the nucleons, in fm),
 r.m.s. radii of the charmed baryon(or bottom), neutron
 and proton, $r_y$, $r_n$ and $r_p$ (in fm), respectively, including the contribution of the $\rho$ mesons.
 The configuration of hyperon is $1s_{1/2}$ for all
hypernuclei. The results of $\Lambda_c^+$ and $\Lambda_b$
hypernuclei are given with $U_{\Lambda_c^+}=U_{\Lambda_b}=-30$
MeV. The results of $\Xi_c$ hypernuclei are given with
$U_{\Xi_c}=-16$ MeV.} \label{t2}
 {\small\begin{center}
 \begin{tabular}{cccccc|ccccccc}  \hline
 $^AZ$ & $-E/A$ &$r_{ch}$&$r_{\rm y}$&$r_n$&$r_p$ &$^A Z$& $-E/A$ &$r_{ch}$&$r_{\rm y}$&$r_n$&$r_p$ \\ \hline
   $^6{\rm Li}$&5.67&2.51&&2.32&2.37 &$^{16}{\rm O}$                     &  8.04 &2.70    & &2.55
   &2.58\\\hline
$^7_{\Lambda_c^+}{\rm
Li}$&5.99&2.42&1.88&2.23&2.28&$^{17}_{\Lambda_c^+}{\rm
O}$&8.33&2.72&2.04&2.56&2.59\\
$^7_{\Lambda_b}{\rm
Li}$&7.04&2.37&1.39&2.19&2.22&$^{17}_{\Lambda_b}{\rm O}$
&8.87&2.71&1.57&2.56&2.58\\
 $^7_{\Xi_c^0}{\rm Li}$& 5.17&2.38&2.59&2.37&2.24&$^{17}_{\Xi_c^0}{\rm O}$&7.97   &2.68   &2.39       & 2.58  &2.55     \\
$^7_{\Xi_c^+}{\rm Li}$&4.90&2.59&2.97&2.22&2.46&$^{17}_{\Xi_c^+}{\rm O}$ &7.71   &2.74    &2.55       &2.53 &2.61 \\
\hline\hline
 $^{10}{\rm B}$&6.22&2.46&&2.29&2.32&$^{40}{\rm Ca}$& 8.52 &3.46 &&3.31 &3.36
 \\\hline
$^{11}_{\Lambda_c^+}{\rm
B}$&6.87&2.43&1.70&2.26&2.29&$^{41}_{\Lambda_c^+}{\rm Ca}$
&8.64&3.47&2.48&3.32&3.37\\
$^{11}_{\Lambda_b}{\rm
B}$&7.86&2.36&1.11&2.19&2.21&$^{41}_{\Lambda_b}{\rm
Ca}$&8.94&3.46&1.94&3.32&3.36\\
    $^{11}_{\Xi_c^0}{\rm B}$&6.14&2.41&2.24&2.33&2.26&$^{41}_{\Xi_c^0}{\rm Ca}$         &8.56  &3.44  &2.70        &3.33   &3.34    \\
  $^{11}_{\Xi_c^+}{\rm B}$&5.86&2.50&2.42&2.25&2.36&$^{41}_{\Xi_c^+}{\rm Ca}$&8.35   &3.48   &2.89       &3.30   &3.38     \\
                      \hline
   \hline$^{12}{\rm C}$                  &7.47    &2.46         && 2.30
&2.32  &$^{208}{\rm Pb}$ &             7.90  &5.51  & &5.71 &5.45
\\\hline
$^{13}_{\Lambda_c^+}{\rm
C}$&8.13&2.43&1.59&2.26&2.29&$^{209}_{\Lambda_c^+}{\rm
Pb}$&7.89&5.51&4.65&5.71&5.45\\
$^{13}_{\Lambda_b}{\rm
C}$&7.90&2.44&2.13&2.28&2.30&$^{209}_{\Lambda_b}{\rm
Pb}$&7.99&5.51&3.64&5.71&5.45\\
     $^{13}_{\Xi_c^0}{\rm C}$         &7.42   &2.41   &2.13       &
2.33  &2.27              &$^{209}_{\Xi_c^0}{\rm Pb}$         &7.90  &5.50  &4.26        &5.72   &5.44    \\
$^{13}_{\Xi_c^+}{\rm C}$        &7.13   &2.49   &2.29 &2.26
&2.35 &$^{209}_{\Xi_c^+}{\rm Pb}$           &  - &-   &   -    & -  &   -  \\

                           \hline\hline

\end{tabular}
 \end{center}}
\end{table}

\newpage

\begin{table}[pthb]
 \caption{Binding energy per baryon, -$E/A$ (in MeV), r.m.s. charge
radius $r_{ch}$(those of the nucleons, in fm),
 r.m.s. radii of the hyperon, neutron
 and proton, $r_y$, $r_n$ and $r_p$ (in fm), respectively, without the contribution of the $\rho$ mesons.
 The configuration of hyperon is $1s_{1/2}$ for all
hypernuclei. The results of $\Xi$ and $\Xi_c$ hypernuclei are
given with $U_\Xi=U_{\Xi_c}=-16$MeV.} \label{t3}
 {\small\begin{center}
 \begin{tabular}{cccccc|ccccccc}  \hline

 $^AZ$ & $-E/A$ &$r_{ch}$&$r_{\rm y}$&$r_n$&$r_p$ &$^A Z$&$-E/A$ &$r_{ch}$&$r_{\rm y}$&$r_n$&$r_p$ \\ \hline
   $^6{\rm Li}$&5.67&2.51&&2.32&2.37 &$^{16}{\rm O}$                     &  8.04 &2.70    & &2.55
   &2.58\\\hline
$^7_{\Xi^-}{\rm Li}$&5.17&2.46&3.07&2.28&2.31&$^{17}_{\Xi^-}{\rm O}$ &8.11   &2.70    &2.58       &2.55 &2.57 \\
 $^7_{\Xi^0}{\rm Li}$& 4.97&2.48&3.38&2.29&2.34&$^{17}_{\Xi^0}{\rm O}$&7.87   &2.71   &2.77       & 2.55  &2.58     \\
$^7_{\Xi_c^0}{\rm Li}$&5.32&2.45&2.21&2.27&2.31&$^{17}_{\Xi^0_c}{\rm O}$ &8.03   &2.71    &2.20       &2.55 &2.58 \\
 $^7_{\Xi^+_c}{\rm Li}$& 5.00&2.48&2.45&2.28&2.34&$^{17}_{\Xi^+_c}{\rm O}$&7.74   &2.71   &2.41       & 2.55  &2.58
 \\\hline
 $^{10}{\rm B}$&6.22&2.46&&2.29&2.32&$^{40}{\rm Ca}$& 8.52 &3.46 &&3.31 &3.36
 \\\hline
  $^{11}_{\Xi^-}{\rm B}$&6.21&2.44&2.54&2.28&2.30&$^{41}_{\Xi^-}{\rm Ca}$&8.73   &3.45   &2.73       &3.31
  &3.35
\\
    $^{11}_{\Xi^0}{\rm B}$&5.97&2.45&2.73&2.29&2.31&$^{41}_{\Xi^0}{\rm Ca}$&8.52  &3.46  &2.96        &3.31   &3.36    \\
  $^{11}_{\Xi^0_c}{\rm B}$&6.26&2.44&1.97&2.28&2.30&$^{41}_{\Xi^0_c}{\rm Ca}$&8.58   &3.46   &2.52       &3.31   &3.36
\\
    $^{11}_{\Xi^+_c}{\rm B}$&5.93&2.45&2.15&2.28&2.31&$^{41}_{\Xi^+_c}{\rm Ca}$&8.35  &3.46  &2.87        &3.31   &3.36    \\

   \hline$^{12}{\rm C}$                  &7.47    &2.46         && 2.30
&2.32  &$^{208}{\rm Pb}$&             7.90  &5.51  &
&5.71&5.45\\\hline
$^{13}_{\Xi^-}{\rm C}$&7.51   &2.45    &2.40 &2.29&2.31 &$^{209}_{\Xi^-}{\rm Pb}$  &8.03   &5.50   &3.56       &5.71   &5.44     \\
$^{13}_{\Xi^0}{\rm C}$ &7.26   &2.45   &2.57       &
2.29  &2.31              &$^{209}_{\Xi^0}{\rm Pb}$  &7.92  &5.51  &4.24        &5.71   &5.45    \\
$^{13}_{\Xi^0_c}{\rm C}$&7.54   &2.44    &1.87 &2.28&2.30 &$^{209}_{\Xi^0_c}{\rm Pb}$  &7.93   &5.51   &3.94       &5.71   &5.45     \\
$^{13}_{\Xi^+_c}{\rm C}$ &7.20   &2.45   &2.04       &
2.29  &2.31              &$^{209}_{\Xi^+_c}{\rm Pb}$  & - &  -& -    & -  & -   \\

                      \hline\hline

\end{tabular}
\end{center}}
\end{table}

\end{document}